\documentclass[fleqn,usenatbib]{mnras}

\usepackage{graphicx}
\usepackage{amsmath}	
\usepackage{amssymb}	
\usepackage{dcolumn}
\usepackage{bm}
\usepackage{xcolor}
\usepackage{algorithm}
\usepackage{algpseudocode}
\usepackage{lastpage}

\title[Growing pains]{Growing Pains: Understanding the Impact of Likelihood Uncertainty on Hierarchical Bayesian Inference for Gravitational-Wave Astronomy}

\author[Colm Talbot et al.]{
Colm Talbot$^{1,2,3}$\thanks{colm.talbot@ligo.org}, Jacob Golomb$^{4,5}$\\
${}^{1}$ LIGO Laboratory, Massachusetts Institute of Technology, 185 Albany St, Cambridge, MA 02139, USA\\
${}^{2}$ Department of Physics and Kavli Institute for Astrophysics and Space Research,\\Massachusetts Institute of Technology, 77 Massachusetts Ave, Cambridge, MA 02139, USA\\
${}^{3}$ Kavli Institute for Cosmological Physics, University of Chicago, 5640 S. Ellis Ave., Chicago, IL 60615, USA\\
${}^{4}$ LIGO Laboratory, California Institute of Technology, Pasadena, CA 91125, USA\\
${}^{5}$ Department of Physics, California Institute of Technology, Pasadena, CA 91125, USA
}

\date{\today}

\begin{document}
\label{firstpage}
\pagerange{\pageref{firstpage}--\pageref{LastPage}}
\maketitle

\begin{abstract}
Observations of gravitational waves emitted by merging compact binaries have provided tantalising hints about stellar astrophysics, cosmology, and fundamental physics.
However, the physical parameters describing the systems, (mass, spin, distance) used to extract these inferences about the Universe are subject to large uncertainties.
The most widely-used method of performing these analyses requires performing many Monte Carlo integrals to marginalise over the uncertainty in the properties of the individual binaries and the survey selection bias.
These Monte Carlo integrals are subject to fundamental statistical uncertainties.
Previous treatments of this statistical uncertainty has focused on ensuring the precision of the inferred inference is unaffected, however, these works have neglected the question of whether sufficient accuracy can also be achieved.
In this work, we provide a practical exploration of the impact of uncertainty in our analyses and provide a suggested framework for verifying that astrophysical inferences made with the gravitational-wave transient catalogue are accurate.
Applying our framework to models used by the LIGO-Virgo-KAGRA collaboration and in the wider literature, we find that Monte Carlo uncertainty in estimating the survey selection bias is the limiting factor in our ability to probe narrow population models and this will rapidly grow more problematic as the size of the observed population increases.
\end{abstract}

\begin{keywords}
gravitational waves — methods: data analysis — methods: statistical
\end{keywords}

\section{\label{sec:introduction}Introduction}

Using data from the first three observing runs of Advanced LIGO~\citep{AdvancedLIGO} and Advanced Virgo~\citep{AdvancedVirgo} $\approx 70$ signals from the merger of compact binary systems have been identified~\citep{GWTC-3}, along with a few tens of less-significant additional candidate events~\citep{4-OGC, Olsen2022}.
While individual observations of compact binary mergers provide insights into astrophysics and cosmology, maximising the physical resolving power using the catalogue of gravitational-wave transients requires analysing the entire population as a hierarchical Bayesian inference problem.
Due to computational constraints, these analyses are performed using a multi-stage process to calculate the population-level likelihood~\cite[see, e.g.,][]{O3bPop, Thrane2019, Vitale2020, Mandel2019}.

First, segments of data that are likely to contain gravitational-wave signals are identified by search pipelines~\citep[e.g.,][]{Allen2012}.
These pipelines are only sensitive to the loudest signals and so the observed sample is biased in favour of nearby high-mass binaries with black hole angular momenta (``spins'') aligned with the orbital angular momentum~\citep{Campanelli2006}.
This selection bias is typically accounted for by estimating the fraction of binaries that we expect to observe using simulated ``injection'' campaigns~\cite{}.

Next, the strain data from gravitational-wave detectors containing the observed transients is analysed with a fiducial reference model for the population (often referred to as the fiducial prior distribution) in order to obtain samples from the fiducial posterior probability distribution for the parameters (masses, spins, etc.) of each binary.
While the fiducial prior distribution impacts the fiducial posterior, it is typically chosen to avoid imprinting astrophysical assumptions on the results.
For example, binaries are assumed to be distributed homogeneously and isotropically throughout the Universe.
The fiducial model for black hole masses is usually uniform in the mass of each black hole and uniform in spin magnitude and isotropic in direction.

In the final stage, these fiducial samples are importance sampled (``reweighted'') using a parameterised model for the underlying population to compute the likelihood for the observed data given population-level parameters (e.g., the maximum allowed black hole mass) marginalised over the per-event parameters.
For each model for the underlying population, the fraction of observable binaries is also estimated using importance sampling on the injected signals from the first stage~\citep[e.g.,][]{Finn1993, Loredo2004, Farr2015}.

The importance sampling step is an example of using Monte Carlo summation to approximate an integral and as such comes with some intrinsic uncertainty that enters the analysis as a source of systematic error.
Typically, this uncertainty is ignored when performing the analysis, however, in recent years several attempts have been made to quantify this uncertainty and theoretically motivated heuristics have been proposed to estimate and (hopefully) mitigate its impact~\citep{Farr2019, Essick2022}.
In this work, we perform a data-driven analysis of the potential systematic uncertainties from our use of Monte Carlo integration.
We note that while we apply our formalism to the problem of population inference for gravitational-wave astronomy, it is widely applicable to any context in which an approximate estimator of the true likelihood is used in a Bayesian analysis.

The remainder of this paper is structured as follows.
In the next Section, we describe how uncertainty appears in our estimate of the population likelihood through Monte Carlo integration and suggest a set of convergence criteria.
In Section~\ref{sec:quantify}, we analyse a simple toy model to examine the impact of uncertainty on the accuracy of inference.
Using this, we establish a threshold beyond which we expect our results to be significantly biased.
Following this, we take a range of models previously considered for population analyses and quantify the uncertainty in these results in Section~\ref{sec:real-data}.
Finally, we provide a closing discussion.

\section{Uncertainty in the Population Likelihood Approximation}\label{sec:uncertainties}

The likelihood function typically employed for an analysis of a population of $N$ observed systems with source-dependent selection effects can be written~\citep[see, e.g.,][for details]{Thrane2019, Vitale2020, Mandel2019}
\begin{equation}\label{eq:log-likelihood}
    {\cal L}(\{d_i\} | \Lambda) \propto \prod^{N}_{i} \frac{{\cal L}(d_i | \Lambda)}{P_{\rm det}(\Lambda)}.
\end{equation}
Here, the $\{d_{i}\}$ are the data containing the observed signals (indexed by $i$).
In the context of gravitational-wave astronomy, this is strain data recorded by gravitational-wave interferometers.
The selection function $P_{\rm det}$ is the fraction of all signals that would be observed for a population described by population hyper-parameters $\Lambda$.
We note that this likelihood has been marginalised over the overall rate of events (assuming a uniform-in-log rate prior) and the parameters describing each of the individual systems.

Each of the terms ${\cal L}(d_{i} | \Lambda)$ and $P_{\rm det}(\Lambda)$ are computed by marginalising over $\theta$, the $\approx 15$ parameters describing the individual binaries and many more describing the noise properties of the interferometers
\begin{align} \label{eq:per-event-likelihood}
    {\cal L}(d_{i} | \Lambda) &= \int d\theta p(d_{i}, \theta | \Lambda) = \int d\theta {\cal L}(d_{i} | \theta) p(\theta | \Lambda) \\
    P_{\rm det}(\Lambda) &= \int dd \int d\theta p(d, \theta | \Lambda) \Theta(\rho(d) - \rho_{*}) \\
    &= \int dd \int d\theta {\cal L}(d | \theta) p(\theta | \Lambda) \Theta(\rho(d) - \rho_{*}).
\end{align}
In both expressions, we have expanded the joint distribution for the observed data and signal parameters into the population model $p(\theta | \Lambda)$ and the likelihood of observing data given single-event parameters ${\cal L}(d | \theta)$.
The integral over $d$ in the expression for $P_{\rm det}$ is over all of the data collected by the instrument while the $d_{i}$ represents the data around the time of a specific observed signal.
The final term is a Heaviside step function for the detection statistic (e.g., signal-to-noise ratio or false-alarm rate) $\rho$ with threshold $\rho_{*}$.
In order to minimise the cost of performing the analysis, these integrals are commonly computed using Monte Carlo estimators using some reference set of samples from the fiducial posterior distribution.
We denote the estimator of quantity $x$ as $\hat{x}$.
As a specific example, the estimator of the log-likelihood (Eq. \ref{eq:log-likelihood}) is
\begin{equation}
    \ln \hat{{\cal L}}(\{d_i\} | \Lambda) = \left(\sum^{N}_{i} \ln \hat{{\cal L}}(d_{i} | \Lambda)\right) - N \ln \hat{P}_{\rm det}(\Lambda).
\end{equation}

In practice, these estimates are calculated using Monte Carlo integration:
\begin{align}
    I &= \int dx f(x) p(x) \equiv \langle f \rangle_{p(x)} \\
    \hat{I} &= \frac{1}{M}\sum^{j=M}_{x_{j} \sim p(x)} f(x_{j}).
\end{align}
Here $\hat{I}$ is the estimator of the integral $I$ and $M$ is the number of samples in the Monte Carlo integral.
We note that $p(x)$ is a normalised probability distribution and $f(x)$ is an arbitrary function of parameters $x$.
Every Monte Carlo has an intrinsic statistical uncertainty
\begin{equation}
    \sigma^{2}_{I} = \frac{1}{M} \left[ \langle f^2 \rangle_{p(x)} - \langle f \rangle^{2}_{p(x)} \right] \equiv \frac{1}{M} \bar{\sigma}^{2}_{I}.
\end{equation}
We define the quantity $\bar{\sigma}^{2}_{I}$ as the intrinsic variance between the proposal distribution $p(x)$ and the target distribution $f(x)p(x)$.
In general, the uncertainty in a Monte Carlo integral will be minimised by choosing $p(x)$ and $f(x)$ to minimise $\bar{\sigma}_{I}$.
For example, for most gravitational-wave population analyses (including this work) we choose
\[
f(\theta) \sim \frac{p(\theta | \Lambda)}{p(\theta | \varnothing)}, \quad
p(\theta) \sim {\cal L}(d | \theta) p(\theta | \varnothing),
\]
where $p(\theta | \varnothing)$ is the fiducial prior distribution.
However, in some cases it is beneficial to define~\citep[e.g.,][]{Wysocki2019, Golomb2022a} $f(\theta) \sim {\cal L}(d | \theta)$, $p(\theta) \sim p(\theta | \Lambda)$.
We also note that the variance scales inversely with the number of samples.
A final quantity related to Monte Carlo integrals that we will need is the effective number of independent samples~\citep{Kish}
\begin{equation}
    n_{\rm eff} = M \frac{\langle f \rangle^{2}_{p(x)}}{\langle f^{2} \rangle_{p(x)}}.
\end{equation}
In~\cite{Farr2019} the author demonstrates that for small values of $n_{\rm eff}$ a Gaussian approximation to the likelihood uncertainty breaks down.
In previous works~\citep[e.g.,][]{Farr2019, O3bPop}, $n_{\rm eff}$ has been used to assess the convergence of the likelihood estimator and to impose data-dependent cuts on the allowed parameter space.
We prefer to work directly with the estimated variance and include $n_{\rm eff}$ here just to compare with previous work.

Since we assume that the reference samples used in each of the Monte Carlo integrals are independent, the variance in the estimate of the population (log-)likelihood is
\begin{equation}
    \sigma^{2}_{\ln \hat{\cal L}}(\Lambda) = \sum^{N}_{i} \sigma^{2}_{\ln \hat{\cal L}_{i}}(\Lambda) + N^2 \sigma^{2}_{\rm sel}(\Lambda) .
\end{equation}
We note that the contribution to the total variance from the selection function grows quadratically with the population size, as ${\rm Var}(N x) = N^2 {\rm Var}(x)$.

Assuming the individual observations are independent and identically distributed draws from the underlying population, we recast this expression in terms of an average per-observation uncertainty $\sigma_{\rm obs}$ to more clearly see the dependence of both terms with the population size
\begin{equation}\label{eq:population-variance}
    \sigma^{2}_{\ln \hat{\cal L}}(\Lambda) = N \sigma^{2}_{\rm obs}(\Lambda) + N^2 \sigma^{2}_{\rm sel}(\Lambda) .
\end{equation}
We have explicitly retained the dependence of this variance on the hyperparameters.
We justify the assumption that $\sigma_{\rm obs}$ does not vary with time in Section~\ref{sec:time-evolution}.

Since we are predominantly interested in differences in log-likelihood for points with significant posterior support, we need to limit the error in the difference of log-likelihood estimators, $\Delta \ln \hat{\cal L}$.
In general, the errors will not be independent, and so we calculate the variance in this quantity $\sigma^{2}_{\Delta \ln \hat{\cal L}}$ as defined in Eq.~A11 in~\cite{Essick2022}.
We assume the error in the estimator of the log-likelihood is Gaussian distributed as the contribution to the population log-likelihood from the per-event terms is the sum of $N$ independently and identically distributed estimators and so by the central limit theorem follows a normal distribution and in the high effective-sample size limit the selection function term also follows a normal distribution~\citep{Farr2019}
We therefore write $\sigma^{2}_{\Delta \ln {\cal L}} = \sigma^{2}_{\Delta \ln \hat{\cal L}}$.

If the uncertainties in the estimators are uncorrelated with $\Lambda$, we will have $\sigma^{2}_{\Delta \ln \hat{\cal L}} = \sigma^{2}_{\ln \hat{\cal L}}$.
In~\cite{Essick2022}, the authors demonstrate that under certain conditions the variance in likelihood differences in ``local neighbourhoods'' avoids the worst-case scaling in Eq.~\ref{eq:population-variance} and rather find that
\begin{equation}
    \sigma^{2}_{\Delta \ln \hat{\cal L}} = \sigma^{2}_{\rm obs}(\Lambda) + N \sigma^{2}_{\rm sel}(\Lambda)
\end{equation}
for a simple example model due to correlation of the Monte Carlo errors between points with significant posterior support.
It is unclear a priori when the local neighbourhood approximation is valid, in this work, we numerically test whether this scaling holds for the specific case of inferring the population properties of merging binary black hole systems.

\subsection{Uncertainty as a draw from a Gaussian process}\label{sec:gp-uncertainty}

To build an understanding of the impact of uncertainty, we assert that the estimated difference in log-likelihood is a fair draw from the Gaussian process with mean function $\Delta \ln {\cal L}$ and (potentially non-stationary) kernel function $\Sigma(\Lambda, \Lambda')$
\begin{equation}
    \Delta \ln \hat{\cal L}(\{d_{i}\} | \Lambda, \Lambda')
    \sim {\cal GP}(\Delta \ln {\cal L}(\{d_{i}\} | \Lambda, \Lambda'), \Sigma(\Lambda, \Lambda')).
\end{equation}
Here $\Sigma(\Lambda, \Lambda') = \sigma^{2}_{\Delta \ln {\cal L}}$ is the $2D$-dimensional covariance matrix, where $D$ is the dimensionality of the population model.
In practice, we do not have access to the true kernel function, and so we approximate it using a numerical covariance matrix using the covariance between the likelihood estimator at each pair of points we consider.
Specifically, we construct the approximate kernel by numerically calculating
\begin{equation}
    \Sigma(\Lambda, \Lambda')
    = \sigma^{2}_{\Delta \ln \hat{\cal L}}
\end{equation}
following Eq.~A11 in~\cite{Essick2022}.
We will use this quantity to estimate the average variance over the posterior for the hyperparameters
\begin{align}
    \left\langle \Delta \ln \hat{\cal L} \right\rangle
    &\equiv \int d\Lambda \int d\Lambda' p(\Lambda | \{d_{i}\}) p(\Lambda' | \{d_{i}\}) \Sigma(\Lambda, \Lambda')
    \\
    &\approx \frac{1}{K^2} \sum^{k=K}_{k=1} \sum^{k'=K}_{k'=1} \Sigma(\Lambda_{k}, \Lambda_{k'})
    \\
    \Lambda_{k} &\sim p(\Lambda | \{d_{i}\}).
\label{eq:effective-variance}
\end{align}
We note that this is the average of the covariance matrix weighted by the posterior support.

This is a slightly different statistic than the one considered in~\citet{Essick2022}, where the authors replace the integral over $\Lambda'$ with a fixed value at the mean of the hyper-posterior \[\bar{\Lambda} = \int d\Lambda p(\Lambda | \{d_{i}\}) \Lambda.\]
While the simpler expression used in~\citet{Essick2022} likely produces comparable results for posterior distributions with Gaussian uncertainties, for posteriors with more complex shapes, e.g., multimodality or curving degeneracies, the mean of the posterior is not in general representative of points with significant posterior support.
In contrast, the full integral over $\Lambda, \Lambda'$ ensures that we accurately represent the variance between all pairs of points with posterior support.

\section{How uncertain can we be?}\label{sec:quantify}

Before turning to real examples, we first motivate an acceptable level of uncertainty in the log-likelihood estimator.
Specifically, we want to know a threshold value of $\langle \Delta \ln \hat{\cal L} \rangle$ above which we expect to see significant biases.
To test this, we consider a simple one-dimensional problem where the true posterior distribution is a unit normal distribution.
To verify that the threshold is independent of the structure of the covariance matrix, we perform this experiment with four analytic kernel functions: a block-diagonal kernel where each block is fully correlated with a random number of blocks, a Mat\'{e}rn kernel with $\nu = 5/2$ and random correlation length, a completely uncorrelated kernel, and a completely correlated kernel.
We find that the result is independent of the kernel choice.

\begin{figure}
    \includegraphics[width=\linewidth]{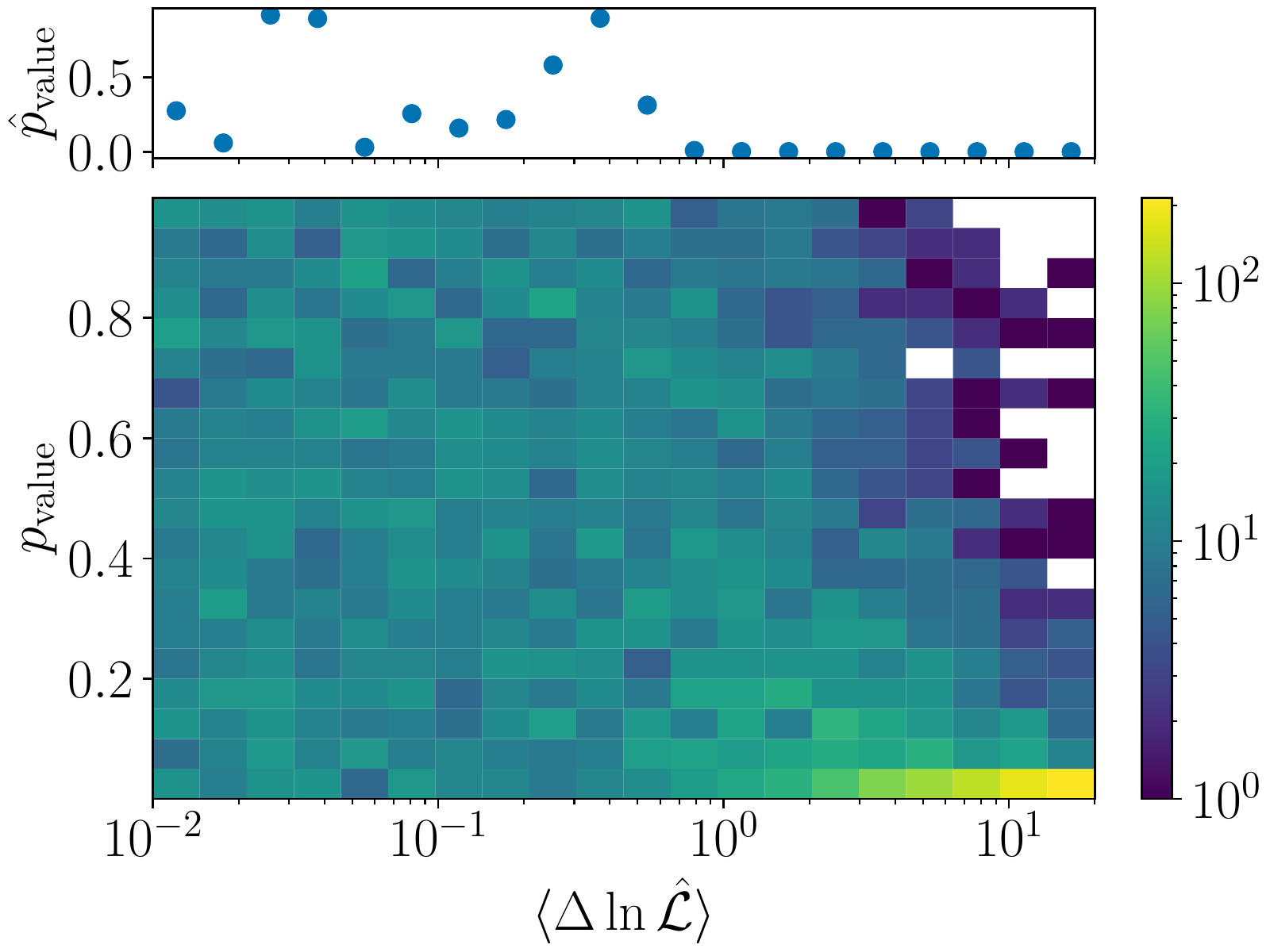}
    \caption{
    $p_{\rm value}$ vs uncertainty in difference in log-likelihood averaged over the posterior distribution ($\langle \Delta \ln \hat{\cal L} \rangle$).
    For unbiased analyses at a given value of $\langle \Delta \ln \hat{\cal L} \rangle$, we expect $p_{\rm value}$ to follow a uniform distribution in $[0, 1]$.
    The upper panel shows a combined $p_{\rm value}$ for all the points in the histogram falling within that range of $\langle \Delta \ln \hat{\cal L} \rangle$.
    We note that this is satisfied for $\langle \Delta \ln \hat{\cal L} \rangle \lesssim 1$, however, when the uncertainty is larger than that value, the analysis is biased on average.
    }
    \label{fig:bias-vs-pvalue}
\end{figure}

For 4800 iterations, we choose a covariance matrix using one of our kernels with a random value of $\langle \Delta \ln \hat{\cal L} \rangle$ drawn logarithmically between $[10^{-2}, 20]$.
For each covariance matrix, we draw 100 realisations from the covariance matrix $\Sigma(\Lambda, \Lambda')$ to generate biased posterior probability distributions.
For each of these realisations, we compute the fraction $f$ of the posterior support below a random point drawn from the true posterior.
If there is no bias, $f$ should follow a uniform distribution in $[0, 1]$.
We, therefore, compute a $p_{\rm value}$ comparing the 100 values of $f$ to the uniform distribution.
In Figure~\ref{fig:bias-vs-pvalue}, we show a two-dimensional histogram of the result of this numerical experiment.
We see that when $\langle \Delta \ln \hat{\cal L} \rangle \lesssim 1$, the $p_{\rm value}$ are uniformly distributed indicating unbiased recovery.
However, as the magnitude of the uncertainty rises, the distribution of $p_{\rm value}$ skews heavily towards small $p_{\rm value}$.
As a final quantitative test, we compare the distribution of $p_{\rm value}$ in each bin of $\langle \Delta \ln \hat{\cal L} \rangle$ to compute a combined $p_{\rm value}$.
We see that the combined $p_{\rm value}$ is very small for $\langle \Delta \ln \hat{\cal L} \rangle \gtrsim 1$.
We will therefore use $\langle \Delta \ln \hat{\cal L} \rangle \gtrsim 1$ as our heuristic threshold for significant bias.

\section{How uncertain are we?}\label{sec:real-data}

We now turn to a tangible example of uncertainty in the inference performed on the population of binary black hole mergers observed during the first three observing runs of Advanced LIGO and Advanced Virgo with a false alarm rate of less than one per year.
The analyses performed in~\cite{O3bPop} imposed cuts on the convergence of the Monte Carlo integrals that implicitly limit the variance in the likelihood to avoid spurious features in the posteriors.
All analyses in that work imposed a condition first proposed in~\cite{Farr2019} demanding that for the selection function $n_{\rm eff} > 4 N$.
Some models also enforced the condition that each marginalization over the single event posteriors had $n_{\rm eff} > N$.
We consider one of the models that applied both convergence conditions.

We compute the uncertainty in the estimated likelihood for one of the models used in the latest LIGO-Virgo-KAGRA analysis.
Specifically, we use the {\tt PowerLaw + Peak} mass model~\citep{Talbot2018}, {\tt Default} spin model~\citep{Talbot2017, Wysocki2019}, and power-law redshift model~\citep{Fishbach2018}.
For our default analysis configuration, we use the same $4278$ per-event posterior samples~\citep{O3bPosteriors} and injection set~\citep{O3bInjections} used in the equivalent analysis in~\cite{O3bPop} and do not apply any constraints on the convergence of the Monte Carlo integrals.

For all of our analyses, we sample the population posterior using the {\tt nestle}~\citep{Nestle} nested sampling package as implemented in {\tt Bilby}~\citep{Bilby}.
We use {\tt GWPopulation}~\citep{Talbot2019} to compute the likelihood function.
We use the same prior distributions as in~\cite{O3bPop}.
For each of the posterior samples, we evaluate the uncertainty in each of the 70 Monte Carlo integrals involved (one for each event and the selection function integral).

\subsection{Evolution of $\sigma_{\rm obs}$}\label{sec:time-evolution}

\begin{figure}
    \centering
    \includegraphics[width=\linewidth]{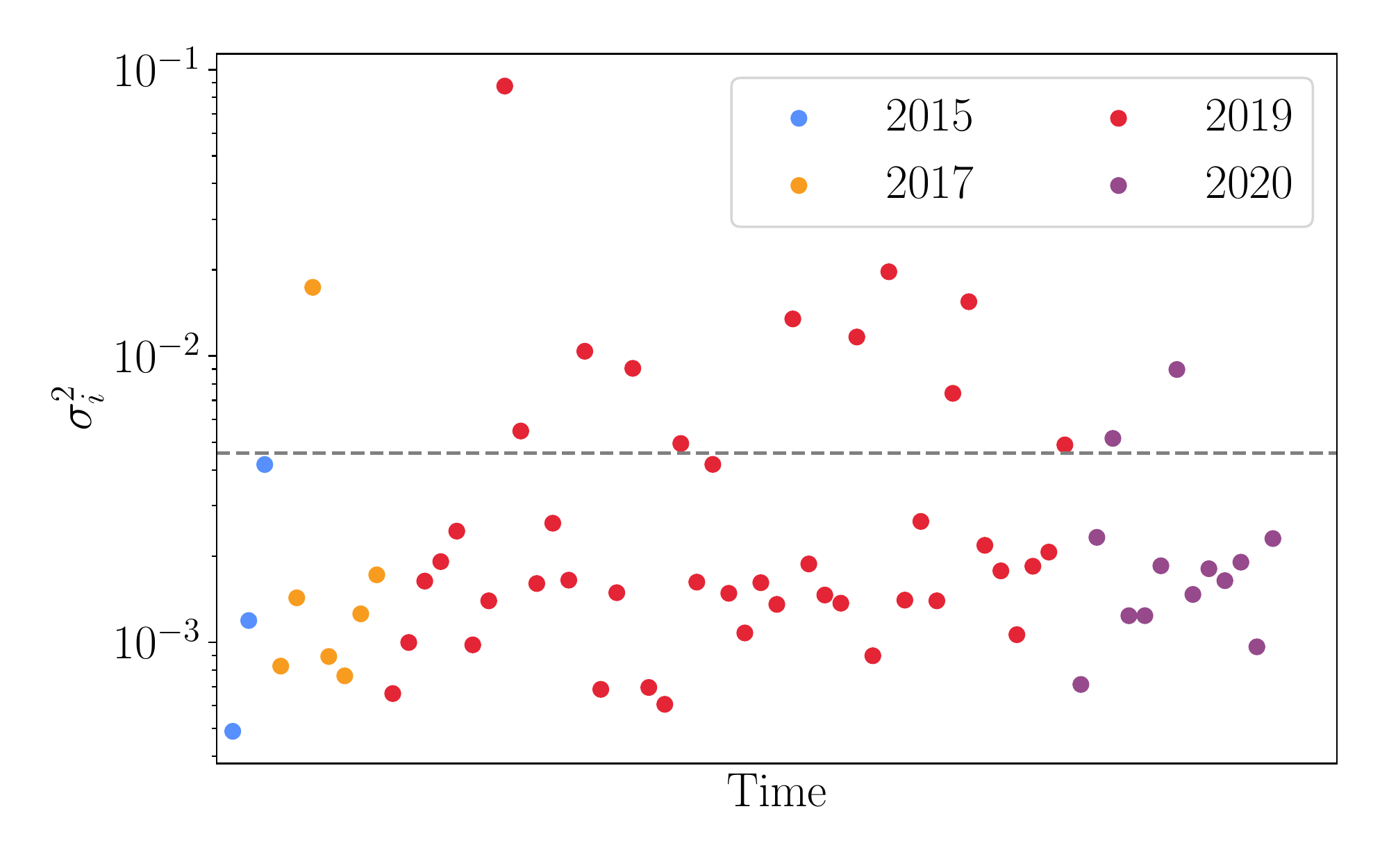}
    \caption{
        The per-event contribution to the likelihood covariance averaged over the posterior support for our population hyperparameters.
        We divide the events by the year of the observation, approximately corresponding to different observing runs of Advanced LIGO/Advanced Virgo.
        We note that there is no obvious trend with time, indicating that we can reliably consider the average uncertainty $\sigma^{2}_{\rm obs} = \langle \sigma^{2}_{i} \rangle$ (shown by the dashed grey line).
    }
    \label{fig:time-evolution}
\end{figure}

We begin by testing our assumption that rewriting the total variance in terms of the average per-event variance $\sigma_{\rm obs}$ is reliable.
One method in which this could break down is if the average uncertainty changes as the sensitivities of the observatories improve.
In Figure~\ref{fig:time-evolution}, we show the average contribution to the covariance over the posterior for the hyperparameters for each event ordered by observation date.
The different colours correspond to events observed in different years.
There is no obvious trend over time which validates our approximation of $\sigma_{\rm obs}^2 = \langle \sigma^{2}_{i} \rangle$.
We show this value with the dashed grey line.
The event with the largest contribution to the uncertainty is GW190517, which has masses consistent with the excess at $\sim 35 M_{\odot}$ and large inferred spins.

\subsection{Scaling with the population size}

\begin{figure}
    \centering
    \includegraphics[width=\linewidth]{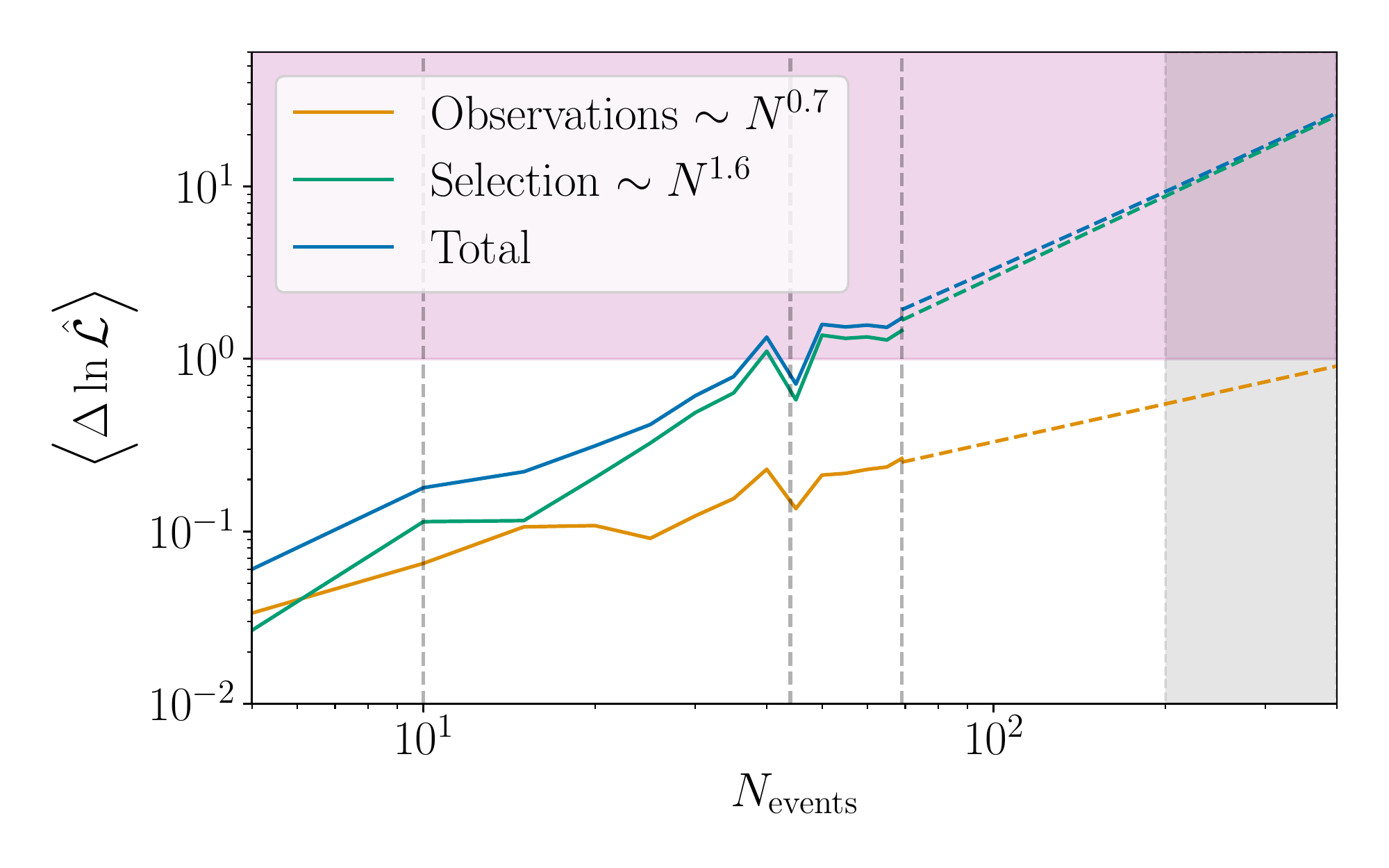}
    \caption{
        Scaling of the uncertainty in the log-likelihood averaged over the full posterior support with the population size for a simple parametric population model.
        The dashed vertical lines show the number of confident binary black hole events in the gravitational-wave transient catalogue at the time of publication of GWTC-1~\citep{GWTC-1}, GWTC-2~\citep{GWTC-2}, and GWTC-3~\citep{GWTC-3}.
        The gray filled region indicates the projected number of binary black hole observations during the next observing run of the international gravitational-wave detector network~\citep{Petrov2022, Weizmann2023}.
        The purple shaded region indicates heuristic values for when the uncertainty in the likelihood is likely to cause noticeable bias in the analysis.
        The solid curves show the empirically obtained uncertainties and the dashed curves are extrapolations based on the power-law fit to the per-event contribution (orange) and the contribution from the selection function (green).
        The total uncertainty is shown in blue.
    }
    \label{fig:uncertainty-vs-nevents}
\end{figure}

\begin{figure}
    \centering
    \includegraphics[width=\linewidth]{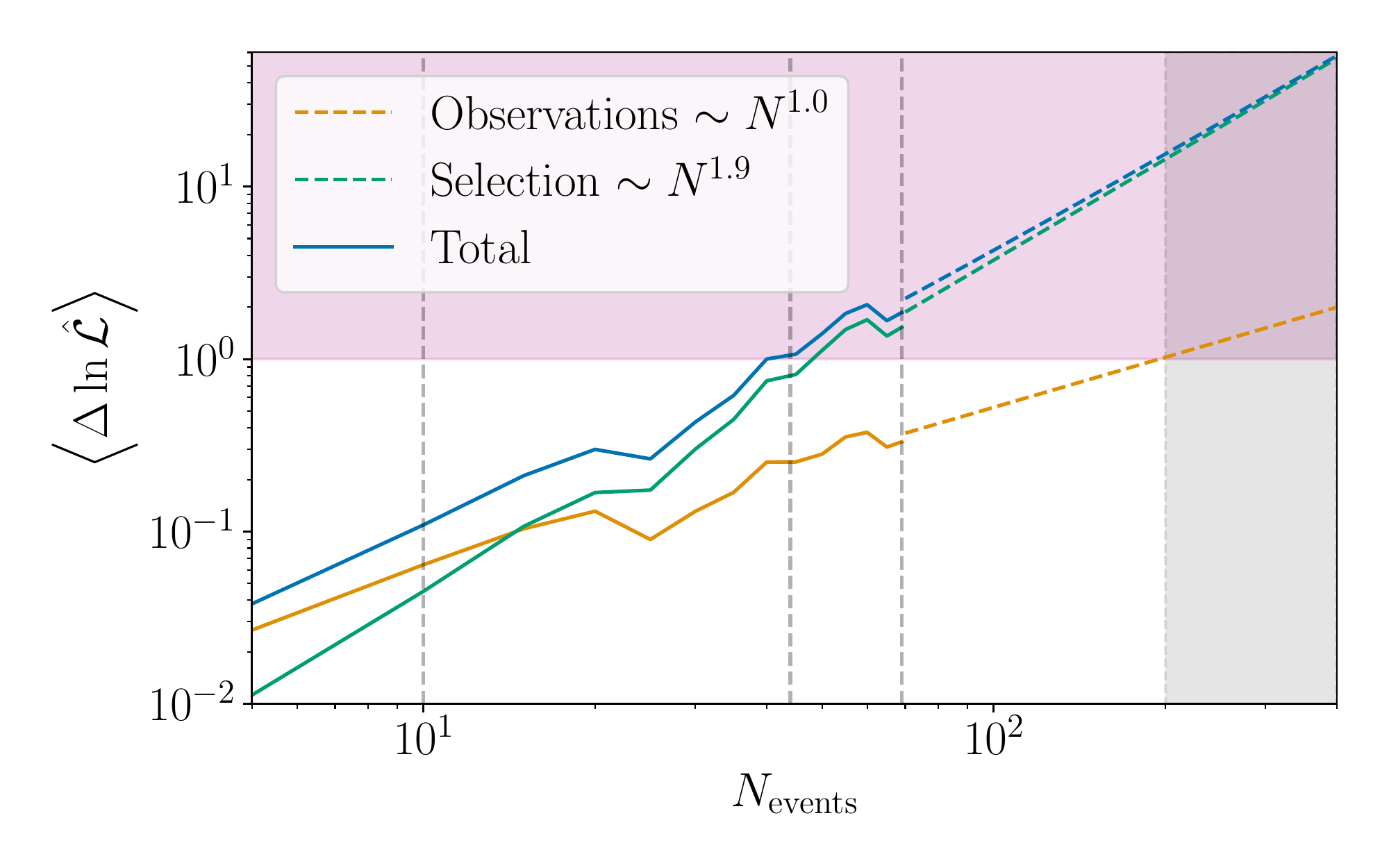}
    \caption{
        The same a Figure~\ref{fig:uncertainty-vs-nevents} but with a more flexible model.
        We note that the same general features are present, however, for this model, the uncertainty grows much more rapidly with population size.
    }
    \label{fig:uncertainty-vs-nevents-spline}
\end{figure}

In order to estimate the scaling of the uncertainty with the size of the catalogue, we randomly sample observations from the total catalogue to simulate smaller catalogues and scale the uncertainty on the selection function appropriately.
Specifically, we consider catalogues with increments of 5 events from 5-65 and all 69 events.
For each catalogue size, we sample from the hyper-posterior and compute the average variance in the estimated differences of log-likelihood values over the posterior samples $\langle \Delta \ln \hat{\cal L} \rangle$.
We do not apply any of the ad-hoc restrictions on Monte Carlo integral convergence proposed in~\cite{Farr2019, O3bPop} and described above in these analyses.
We fit a simple model to the uncertainty coming from the per-event terms and the selection function to obtain fits for the contribution from the individual events and the sensitivity.
The model for the total variance is
\begin{equation}\label{eq:empirical-scaling}
    \langle \Delta \ln \hat{\cal L} \rangle
    = \sigma^2_{\rm obs} N^{a} + \sigma^2_{\rm sel} N^{b}.
\end{equation}
Here, we emphasise that $\Delta \ln {\cal L}$ is proportional to the variance in the estimator and not the standard deviation.
We note that Eq.~\ref{eq:population-variance} implies $a=1$, $b=2$ while if the assumptions from~\cite{Essick2022} hold we will have $a=0$, $b=1$.
We perform this calculation for both the mean variance and the mean covariance over the posterior support.

In Figure~\ref{fig:uncertainty-vs-nevents}, we show the total uncertainty (blue) along with the contributions from the per-event terms (orange) and the selection function (green) as a function of the number of events with the solid curves.
The dashed-coloured curves show projections for larger populations based on the analytic fit.
The dashed grey lines indicate the number of events in each of the first three gravitational-wave transient catalogues and the grey-shaded region shows a plausible range of observations we may expect after the upcoming fourth gravitational-wave observing run~\citep{Petrov2022, Weizmann2023}.
The purple-shaded region shows where, heuristically, we may expect to see noticeable biases, following the criteria developed in Section~\ref{sec:quantify}. 

We find that in practice, the scaling of the uncertainty lies between the best-case scenario from~\cite{Farr2019, Essick2022} and the worst-case scenario in Eq.~\ref{eq:population-variance}.
Specifically, we find $a=0.7$, $b=1.6$.
The dominant source of uncertainty is from estimating the selection function when the population is larger than $\approx 10$ events.
We note that for populations larger than $\approx 40$ events, the uncertainty is consistently in the purple-shaded region.
This is consistent with the fact that ad-hoc cuts on the prior space or Monte Carlo convergence were needed to avoid significant biases in~\cite{O3aPop}.

To test if this scaling depends on the functional form used to fit the population, we repeat the above calculation with a more flexible model for the primary mass and spin parameters.
Specifically, we take the exponential-spline-modulated power-law mass distribution from~\cite{Edelman2022} and the exponential-spline model for black hole spin magnitudes and tilt angles from~\cite{Golomb2023}.
For the mass distribution, we use ten spline nodes spaced logarithmically between $[2, 100] M_{\odot}$ and for the spin parameters we take six nodes equally spaced over the relevant domain.
For all spline nodes, our prior on the amplitudes is a unit normal distribution, except for the endpoints for the mass distribution which are fixed to zero.

In Figure~\ref{fig:uncertainty-vs-nevents-spline}, we show the same as Figure~\ref{fig:uncertainty-vs-nevents} with this more flexible model.
We see that the average covariance in both the per-event and selection function terms grows more rapidly in this case than for the simpler model (a=1.0, b=1.9).
The more extreme scaling may be due to the greater flexibility of the spline model causing the ``local neighbourhood'' assumption of~\cite{Essick2022} to be less appropriate.

\subsection{Scaling with the size of Monte Carlo integrals}

\begin{figure}
    \centering
    \includegraphics[width=\linewidth]{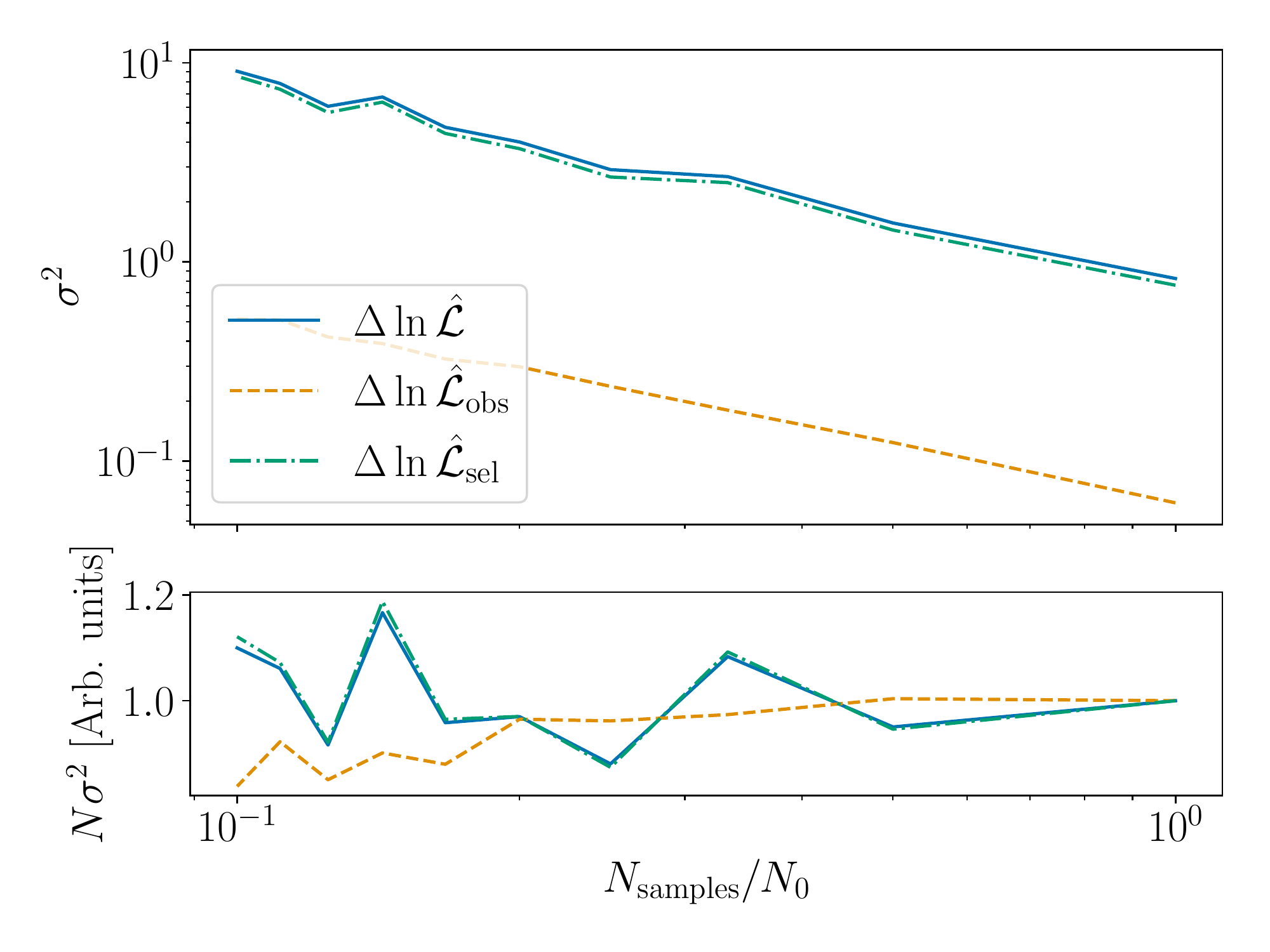}
    \caption{
        The scaling of the average variance in the log-likelihood with the number of events per Monte Carlo integral.
        The solid blue, dashed orange, and dash-dotted green curves show the results using the full likelihood, selection function only, and per-observation terms respectively.
        In the top panel, we show the variance.
        In the bottom panel, we show the normalized variance divided the number of samples per integral.
        As expected, these quantities scale inversely with the number of samples.
    }
    \label{fig:factored-uncertainty-vs-n-samples}
\end{figure}

Having established numerically how the size of the uncertainty in the likelihood estimates varies with the size of the population and configuration settings, we turn to how the number of samples per Monte Carlo integral impacts the uncertainty for this concrete example.
To address this, we repeat the uncertainty calculation for the {\tt PowerLaw + Peak} and  {\tt Default} configuration above ten times, once using all of the available samples, once with half of the samples, one third of the samples, etc., down to one tenth of the samples.

In Figure~\ref{fig:factored-uncertainty-vs-n-samples}, we show the mean variance over the posterior distribution as a function of the number of samples per Monte Carlo integral in the upper panel.
The solid blue, dashed orange, and dash-dotted green curves show the results using the full likelihood, selection function only, and per-observation terms respectively.
In the lower panel, we show the variance scaled by the number of samples in the integral such that it will be constant if the uncertainty scales linearly with the number of samples.
We observe that the variance is consistent with scaling inversely with the number of samples.

\subsection{Impact on the inferred astrophysical distributions}

\begin{table*}
    \begin{tabular}{ |c|c|c|c|c|c|c|c|c|c| } 
        \hline
        & $N_{\rm injections}$ & $\alpha$ & $m_{\max}$ & $m_{\min}$ & $\delta_{m}$ & $\mu_{m}$ & $\sigma_{m}$ & $\lambda$ & $\langle \Delta \ln \hat{\cal L} \rangle$ \\ \hline
        LVK & $4 \times 10^{4}$ & 2 & 100 & 2 & 0 & - & - & 0 & 0.63 \\
        No Convergence & $4 \times 10^{4}$ & 2 & 100 & 2 & 0 & - & - & 0 & 5.06 \\
        Tailored & $4 \times 10^{4}$ & 3.5 & 105 & 3 & 6 & 33 & 5 & 0.04 & 1.24 \\
        More Injections & $8 \times 10^{5}$ & 1 & 100 & 2 & 0 & - & - & 0 & 0.50 \\
        No Injections & 0 & - & - & - & - & - & - & - & 0.42 \\
        \hline
    \end{tabular}
    \caption{
        Hyperparameters for the injection sets used in each of the analysis configurations we consider as described in Section~\ref{sec:real-data}.
        We additionally list the average variance in the difference between estimated likelihood values.
    }
    \label{table:injection-parameters}
\end{table*}

\begin{figure}
    \centering
    \includegraphics[width=\linewidth]{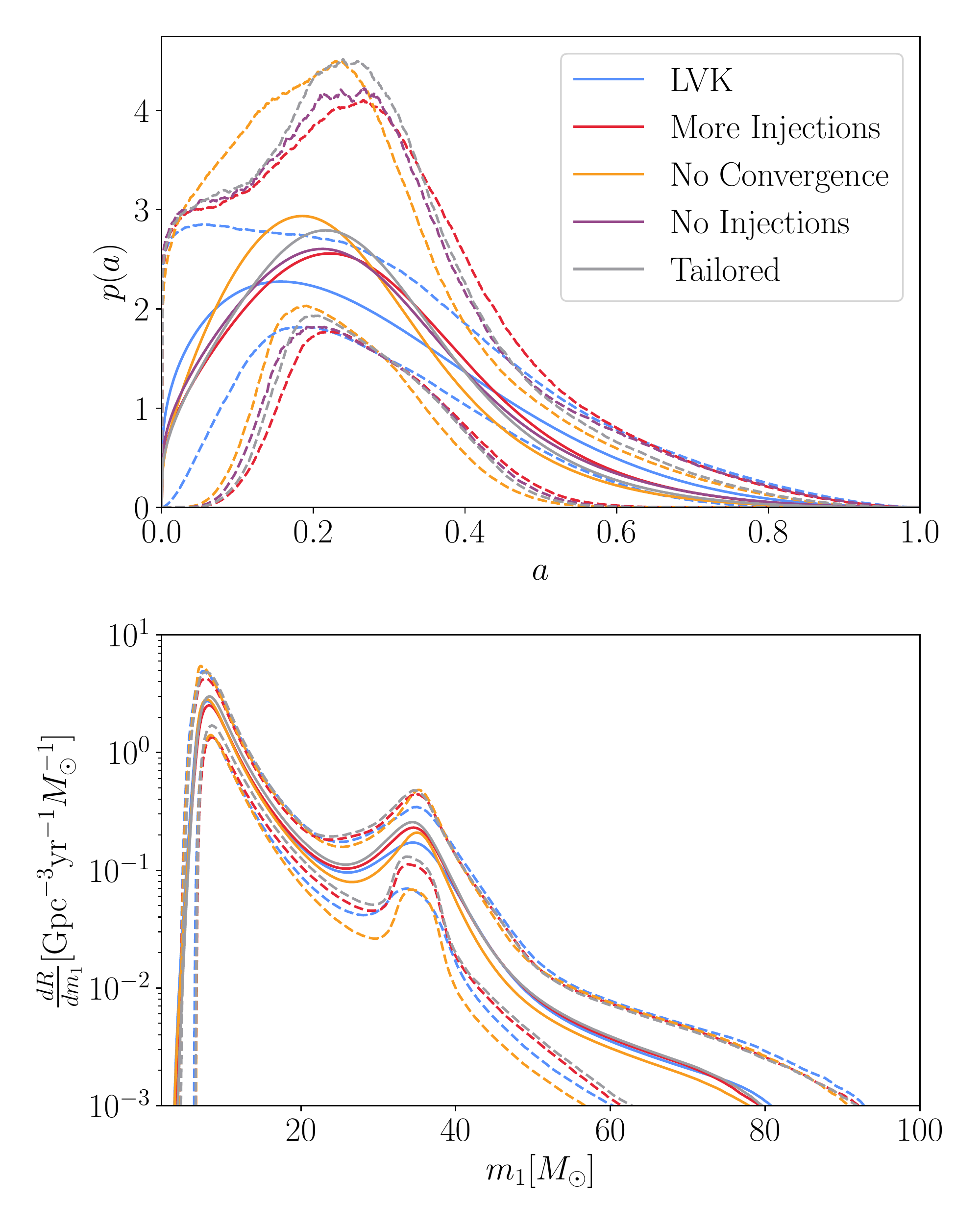}
    \caption{
        The inferred spin magnitude (top) and primary mass (bottom) distributions for a range of analysis configurations.
        The solid curves show the mean inferred distribution and the shaded regions show the $90\%$ symmetric credible interval.
        The blue curves show the results presented in~\protect\cite{O3bPop}.
        In orange, we show results obtained using the same input samples but without performing the ad-hoc constraints on the number of effective samples per Monte Carlo integral.
        In red, we show the results when using more found injections to compute the selection function.
        In purple, we show the results obtained when neglecting the selection function, we note that in this case, we do not show the inferred mass distribution as that is significantly biased by neglecting selection effects.
        In grey, we show the results obtained using our tailored injection set.
    }
    \label{fig:inferred-spectrum-comparisons}
\end{figure}

To study the impact of the convergence-motivated prior cuts and bias in likelihood estimates we consider four analysis configurations:
\begin{itemize}
    \item {\em LVK.} The first configuration is the same one used in the LIGO-Virgo-KAGRA analysis in~\cite{O3bPop}.
    This analysis used $\sim 4\times 10^4$ found injections to estimate the selection function, and $4278$ fiducial posterior samples were used for each event. Specifically, we use the posterior samples released in~\cite{O3bPosteriors} and the set of sensitivity injections which combine injections covering the first three observing runs of Advanced LIGO/Advanced Virgo~\citep{O3bInjections}.
    For this configuration, there is the prior cut on $n_{\rm eff}$ for each of the Monte Carlo integrals as described at the beginning of this section.
    \item {\em No Convergence.} The second configuration repeats the analysis from~\cite{O3bPop} but removes the prior constraints on $n_{\rm eff}$ in each Monte Carlo integral.
    \item {\em Tailored Injections.} We replace the injection set released by the LVK, we use synthetic injections drawn using a mass distribution that more closely matches the observed distribution.
    Specifically, we set the mass distribution using the {\tt PowerLaw+Peak} model using the parameters in Table~\ref{table:injection-parameters}.
    Since the proposal distribution for our Monte Carlo integral more closely matches the target distribution, we expect this injection set to lead to smaller uncertainties with the same number of found injections.
    \item {\em More Injections.} Rather than using the $\sim 4 \times 10^4$ found injections used in~\cite{O3bPop}, we use the $\sim 8 \times 10^5$ synthetic found injections used in~\cite{Golomb2023} in order to reduce the uncertainty in the estimate of the selection function.
    While this uses many more injections, we note that the underlying distribution of signals is different than for the {\em LVK} configuration.
    \item {\em No Injections.} Rather than using the $\sim 4 \times 10^4$ found injections used in~\cite{O3bPop}, we ignore the impact of selection effects completely.
    This will reduce the uncertainty in the estimated likelihoods at the cost of only estimating the observed distribution and not the underlying astrophysical distribution.
\end{itemize}

For both cases where we use synthetic injection sets, we do not repeat the full injection and recovery using a matched-filter search pipeline due to the large associated computational cost.
Instead, we threshold the simulated signals on the optimal signal-to-noise ratio of the injected signal in Gaussian noise with PSDs matching the detector sensitivity during O3 rather than the false-alarm rate~\citep{GWTC-3, GWTC3_data}.
We anticipate that this difference between the detection thresholds does not significantly bias the inferred mass and spin distributions~\citep[e.g.,][]{O2Pop, O3aPop, O3bPop, Golomb2023, Essick2023}.

In Table~\ref{table:injection-parameters} we summarise the population hyper-parameters describing the mass distribution used for each injection set.
Additionally, we show $\langle \Delta \ln \hat{\cal L} \rangle$ computed over the respective posterior distributions for the hyperparameters.
We find that the {\em No Convergence} case clearly surpasses our threshold.
The {\em Tailored} injection set reduces the variance by $\approx 4\times$ by more closely matching the true underlying distribution but is still in the regime where we expect to see some bias.
For the other cases $\langle \Delta \ln \hat{\cal L} \rangle < 1$ and so we would expect the results to be unimpacted by Monte Carlo convergence.

In Figure~\ref{fig:inferred-spectrum-comparisons} we show the inferred distribution for spin magnitude (top panel) and primary mass (lower panel) with our five analysis configurations.
We note that the {\em No Injections} configuration is excluded for the primary mass distribution as that distribution is strongly biased by not accounting for selection effects.
The solid lines indicate the mean inferred distributions and the dashed curves enclose the $90\%$ uncertainty region.
While the uncertainties of all of the results agree within their error bars there are visible differences between the inferred results.
Specifically, we find that for both parameters, the width of the peak at $a \approx 0.2$ and $m_1 \approx 35 M_{\odot}$ are broadest for the result that imposes cuts on the prior based on Monte Carlo convergence (blue) and narrowest for the analysis that has the largest average uncertainties (orange) with the analyses with reduced uncertainty (grey, red, purple) lying in between.
This indicates that for commonly used analysis configurations, the inferred shape of features in the distribution of black hole mass and spin are notably impacted by uncertainty in the estimate of the likelihood.

We note that the inferred spin magnitude distributions for the {\em More Injections} and {\em No Injections} configurations are the most consistent.
This would be the expected outcome if the impact of the spin magnitude on the selection function is small and the uncertainty in the likelihood estimates is small.
We thus infer that the larger injection set is sufficient to remove the bias present when using the found injections released by the LIGO-Virgo-KAGRA collaboration.
While the cuts on the number of effective samples in each Monte Carlo integral in the {\em LVK} configuration control the average uncertainty in the likelihood estimates, the cuts have a visible impact on the inferred distributions.

\subsection{Result differences are explainable due to Monte Carlo uncertainty}

The posterior predictive distribution (PPD) for the binary parameters is defined as
\begin{align}
    p(\theta | \{d\}) = \int d\Lambda p(\theta | \Lambda) p(\Lambda | \{d\}) 
    \approx \frac{1}{N} \sum^{N}_{\Lambda_{i} \sim p(\Lambda | \{d\})} p(\theta | \Lambda_{i}).
\end{align}
In Fig.~\ref{fig:inferred-spectrum-comparisons} the solid curves show the PPD using our different analysis setups (solid curves).
While the curves are visibly different, we wish to know whether the differences can be explained as the result of statistical fluctuations expected due to the uncertainty in our estimator of the likelihood.

\begin{figure}
    \centering
    \includegraphics[width=\linewidth]{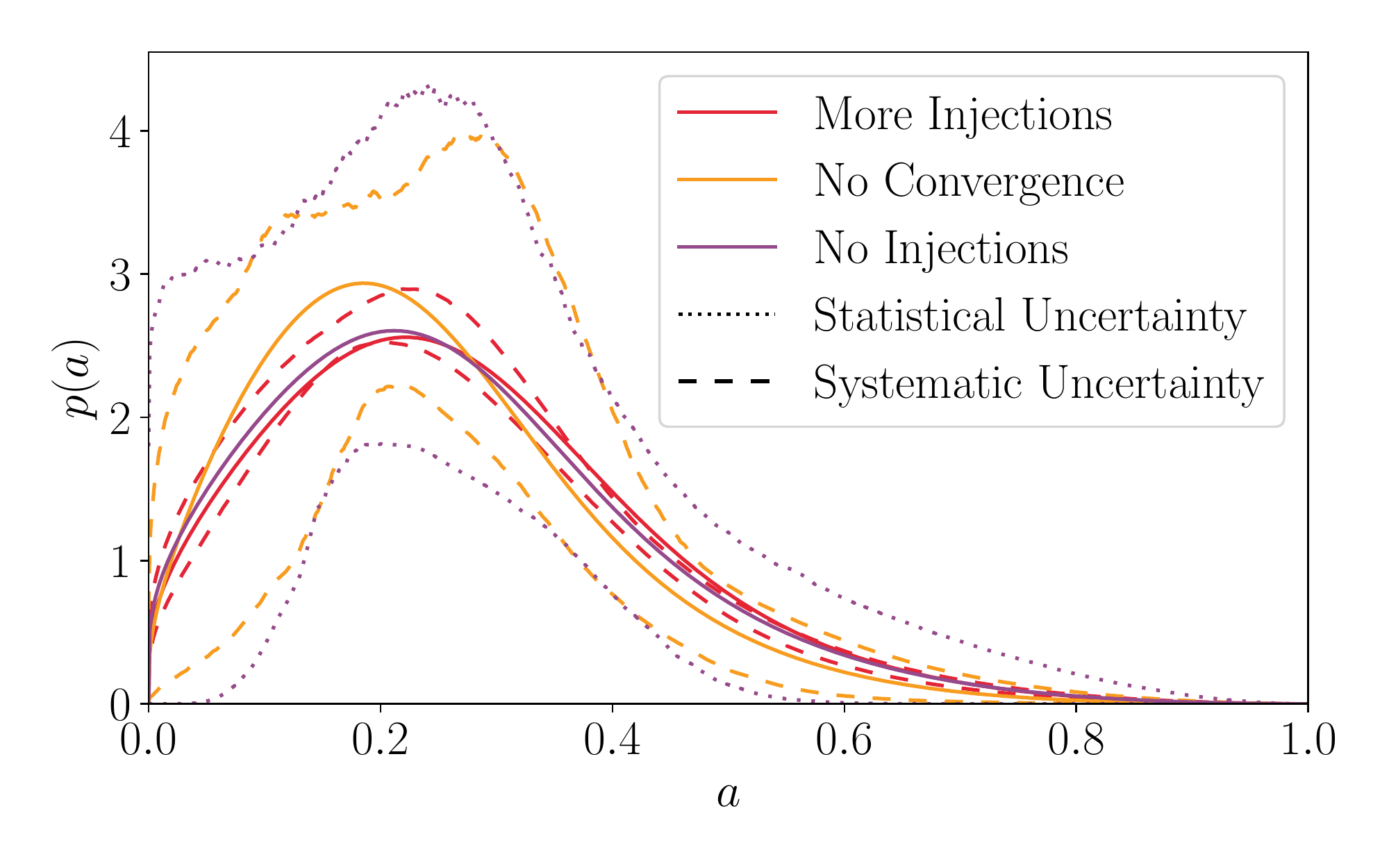}
    \caption{
    Comparison of statistical and systematic uncertainty in our inference of the distribution of black hole spin magnitudes $a$.
    The solid curves show the posterior predictive distribution for three of the analysis configurations described in Section~\ref{sec:real-data}.
    The dotted curves show the 5th and 95th percentiles of our statistical uncertainty for the lowest variance analysis ({\em No Injections}).
    The orange and green dashed curves show the 5th and 95th percentiles of the additional systematic uncertainty from estimating the selection function.
    We note that for the {\em More Injections} case the systematic uncertainty is much smaller than the statistical.
    However, for the {\em No Convergence} case the systematic uncertainty is comparable to the statistical.
    }
    \label{fig:jittered-predictions}
\end{figure}

Our aim is to estimate the range of different PPDs we might expect to measure given the PPD with no systematic uncertainty and a covariance $\Sigma(\Lambda, \Lambda')$.
In the absence of a ground truth, we take the {\em No Injections} case as our reference analysis as it has the lowest uncertainty estimator of the likelihood and neglect the impact of the per-event integrals as all analyses use the same set of samples for each event.

We begin by taking the samples $\Lambda_{i} \sim p(\Lambda | \{d\})$ for the reference case.
We then construct the covariance matrix by numerically calculating the covariance between the likelihood estimates for every pair of posterior samples.
Using this covariance matrix, we generate weights for each of the samples $\delta \sim {\cal N}(0, \Sigma(\Lambda, \Lambda'))$.
Finally, we compute the PPD using these weights as
\begin{align}
    \hat{p}(\theta | \{d\}) = \frac{\langle \delta(\Lambda_{i}) p(\theta | \Lambda_{i}) \rangle_{\Lambda_{i} \sim p(\Lambda | \{d\})}}{\langle \delta(\Lambda_{i}) \rangle_{\Lambda_{i} \sim p(\Lambda | \{d\})}}.
\end{align}
By repeating this many times, we can construct the 90\% credible interval for the systematic error.

In Fig.~\ref{fig:jittered-predictions} we show the same PPDs for the {\em No Injections}, {\em No Convergence}, and {\em More Injections} configurations and the statistical uncertainty for the {\em No injections} configuration (dotted curves) as in Fig.~\ref{fig:inferred-spectrum-comparisons}.
The estimated systematic uncertainty is shown by the dashed curves.
We note that in both cases, the PPDs with our specific realisation are entirely consistent with the systematic uncertainty region indicating that the differences in the PPDs can be fully explained by Monte Carlo uncertainty.
For the {\em No Convergence} case, the estimated systematic uncertainty is comparable to the statistical uncertainty.
One limitation of our method is that the realisations cannot deviate outside the set of samples used for importance sampling and so cannot accurately resolve cases where the systematic uncertainty is larger than the statistical uncertainty in the posterior.

\section{\label{sec:conclusion}Conclusions}

Often when performing Bayesian inference, we cannot calculate the true likelihood function, but rather a computationally tractable approximation.
For example, the use of Monte Carlo integration to approximate marginal likelihoods is widespread in population inference in gravitational-wave astronomy and beyond.
However, often, the uncertainty associated with these finite numerical integrals is neglected.
We specifically examine the requirement of performing unbiased population inference on binary black holes with Monte Carlo integrals used to marginalise over the parameters of the individual sources.
Previous work has claimed that as the size of the population increases, keeping the allowed uncertainty in each marginal likelihood constant (e.g., the number of samples used in each Monte Carlo integral doesn't have to increase with the population size) is sufficient for precise inference of the population parameters~\citep{Essick2022}.

In this work, through a series of numerical experiments, we demonstrated that for models widely used to characterise the population of merging black hole binaries, this scaling is insufficient and the actual scaling depends on the functional form chosen to fit the distribution.
Failing to use a larger number of samples per Monte Carlo integral will result in an increasingly significant bias in the recovery of the population as the number of observations grows.
We recommend that the calculations described in this work be routinely performed for any population analysis to identify cases where the inference may be impacted by Monte Carlo uncertainty.
We provide scripts to evaluate this in the accompanying code release.

By considering a model routinely employed to characterise the distribution of masses and spins of merging compact binaries, we found that the uncertainty in the likelihoods estimated as part of population inference on the gravitational-wave transient catalogue is sufficient to lead to noticeable bias with the current size of the gravitational-wave transient catalogue.
Additionally, by examining the impact of the specific choice of input samples and convergence requirements we observed changes in the width of features in the distribution of black hole masses and spin magnitudes.
While the differences observed here are within the statistical uncertainties, more significant biases have been observed when using more flexible models, e.g., Appendix B of~\cite{Golomb2023}.

The results presented in this work are somewhat in conflict with the results from~\cite{Essick2022}.
One difference between this work and theirs is that in~\cite{Essick2022} the authors only consider population models where the uncertainties on each measurement are smaller than the width of the population.
By contrast, in many of the models considered here, including the models for the black hole mass and spin, the individual measurements are broader than the underlying population model.
The spin magnitudes of individual black holes are very poorly measured, and so the individual posterior distributions are inevitably broader than the population for the majority of systems.
For black hole masses, one might think that the total population model is broader than individual measurements; very few black holes are consistent with masses ranging from $5-80 M_{\odot}$.
However, the relevant quantity is not the whole domain of the model, but rather than change in the population model over the individual event posterior support.
For events intersecting the Gaussian peak at $\sim 35 M_{\odot}$ the uncertainty in the mass is almost always larger than the preferred width of $1-5 M_{\odot}$.
We defer detailed investigations into whether this is a relevant difference to future work.

In the next observing runs, we can conservatively expect the size of the observed population to double or triple~\citep{Petrov2022}.
With a population of this size, we can expect that if we continue to use the same number of samples per Monte Carlo integral the variance in the log-likelihood will reach $\sim 4-10$ and we will be in danger of making severely biased inferences.
In order to avoid this, we will either need to use dramatically more samples in our Monte Carlo integrals or consider novel approaches.

There are a number of questions posed by our results that should be explored in future work.
In Section~\ref{sec:real-data}, we found an approximate scaling for the growth of the uncertainty with the population size, developing a theoretical understanding of this scaling may prove instructive in developing improved methods to deal with large populations.
Ensuring accurate estimation of the population likelihood is an increasingly complex task as the population size increases, and so we will require increasingly sophisticated methods.

As shown in Section~\ref{sec:real-data}, a simple method to reduce the uncertainty in Monte Carlo integrals is to reduce the divergence between the initial model and the target model.
Fortunately, as the size of the population grows, we can use our existing knowledge to generate initial models that well approximate the true distribution, e.g., by drawing our injections to determine the survey sensitivity by our best estimate of the true population.
Additionally, one can recast the Monte Carlo integral using continuous representations of the per-event likelihoods in order to minimise the uncertainty, e.g.,~\cite{Wysocki2019, Golomb2022a}.
Finally, one can limit the analysis to only consider slowly varying source models, e.g., by imposing smoothing priors on the population model~\citep{Edelman2023, Callister2023}.
However, this can lead to missing any sharp features in the distribution.

Each of these improvements is likely to fail eventually, and new methods will be needed.
One possibility is to remove the Monte Carlo integral to determine the selection function and instead directly model the observed distribution of compact binaries.
If desired, the astrophysical distribution can then be obtained as a post-processing stage using continuous estimates of the selection function such as those in, e.g.,~\cite{Veske2021, Talbot2022}.
Similar approaches have been proposed for analyses of online polling data~\citep[e.g.,][]{Elliott2017, Liu2022}.
Since the contribution of the uncertainty from estimating the selection function grows most rapidly with population size, this will significantly alleviate bias in the inferred distribution.

While we considered uncertainties in the likelihood function used for gravitational-wave population inference, our analysis holds for any problem where there are parameter-dependent biases in calculating likelihoods.
For example, when characterising individual compact binary coalescences, there are a number of sources of bias in the likelihood function, including waveform systematics~\citep{Purrer2020}, detector calibration uncertainty~\citep{Payne2020, Vitale2021}, and likelihood acceleration methods~\citep{Smith2016, Morisaki2021, Leslie2021}.
While the specific results in Section~\ref{sec:real-data} will not be relevant to these cases, the general expressions in Sections~\ref{sec:gp-uncertainty} and~\ref{sec:quantify} are relevant.

\section*{acknowledgements}

We thank Reed Essick and Will Farr for multiple discussions.
We thank Sylvia Biscoveanu, Tom Callister, Jack Heinzel, Eric Thrane, and Salvatore Vitale for useful conversations and comments.
JG acknowledges funding from NSF grants 2207758 and PHY-1764464.
CT is supported by an MKI Kavli Fellowship and an Eric and Wendy Schmidt Fellowship for AI in Science at the University of Chicago.
This material is based upon work supported by NSF's LIGO Laboratory which is a major facility fully funded by the National Science Foundation.
This work used computational resources provided by the Caltech LIGO Lab and supported by NSF grants PHY-0757058 and PHY-0823459.
This work made use of the following software: {\tt numpy}~\citep{Oliphant2006, Harris2020}, {\tt cupy}~\cite{Okuta2017}, {\tt nestle}~\citep{Nestle}, {\tt Bilby}~\citep{Bilby}, {\tt GWPopulation}~\citep{Talbot2019}, {\tt gwpopulation\_pipe}~\citep{gwpop_pipe}.

\section*{data availability}
This work used publicly available data produced by the LIGO-Virgo-KAGRA collaborations~\citep{O3bPosteriors, O3bInjections}.
We additionally used a larger synthetic injection set that we will make available on request.
Scripts and {\tt Jupyter} notebooks required to reproduce this analysis are available at \href{https://github.com/ColmTalbot/monte-carlo-uncertainty-scaling}{github.com/ColmTalbot/monte-carlo-uncertainty-scaling}.

\bibliographystyle{mnras}
\bibliography{refs}

\end{document}